\documentclass{article}
\usepackage[utf8]{inputenc}
\usepackage{arabtex}
\usepackage{textcomp}
\usepackage[numbers]{natbib}
\usepackage{mathtools}
\usepackage{float}
\title{Enhanced terahertz radiation from an elliptical-beam-illuminated sawtooth photoconductive antenna: design and numerical analysis}
\author{Jitao Zhang $^{\dag}$\\ ECE Department,The University of Arizona, Tucson, AZ,85721\\
$^{\dag}$ \textit{jitaozhang@email.arizona.edu}}

\date{%
    \today
    \\[2\baselineskip]
    \normalfont\normalsize%
    \parbox{0.8\linewidth}{%
{\bfseries Abstract}: The enhanced terahertz radiation from a new photoconductive antenna (PCA) has been predicted by numerical simulation. Different from the conventional PCA, the proposed PCA has electrodes with sawtooth structures on the edge, which will introduce stronger localized bias field than common electrode (e.g. strip-line structure). In addition, the elliptical beam of the laser source is used to illuminate the sawtooth structure of the PCA, so that the effective region of the terahertz excitation is enlarged and higher laser power can be applied before saturation effect occurs. The design of the proposed PCA is presented, and its performance is predicated by numerical analysis based on the full-wave finite-difference-time-domain method. The simulated result shows that, comparing with a conventional PCA, the proposed PCA achieves $1.4$ times enhancement of the THz radiation field(i.e. peak of the time-domain THz pulse) at the same condition. Further enhancement as high as $2$ times can be achieved when considering the upper limit of the power density of the incoming laser beam.
    }
}
\usepackage{natbib}
\usepackage{graphicx}

\begin{document}

\maketitle

\section{Introduction}

As one of the most commonly used terahertz(THz) source, photoconductive antenna (PCA) has been widely used in THz spectroscopy (such as time-domain spectroscopy) and imaging due to its simple configuration and broadband nature. Since its appearance in 1980s\citep{auston1984}, many efforts have been devoted to improve the performance of the PCA. To date, the biggest issue that hindering the application of the PCA is the ultra-low radiation power\citep{tonouchi2007}. The generation of the THz pulse from a PCA is the result of the instantaneous current excited by the ultra-fast laser pulse (usually in sub-picoseconds). The low radiation power of a PCA is revealed by the fact of the inefficient optics-to-THz conversion, which is usually less than $1\%$\citep{suen2010}. It is further found that the ultra-low quantum efficiency of a PCA ultimately restrict the radiation power of the THz pulse. When the laser pulse illuminates the active region of the PCA, each photon can generate a pair of hole and electron in the active layer in ideal case. The migration of the carriers (In most case, only electron is under consideration because its mobility is much larger than that of hole) driven by the biased field will generate transient current, which will eventually result in THz radiation in free space with the help of the electrode. However, due to the short lifetime, most of the photo-excited electrons are recombined before they can reach the electrode, and thus have little contribution to the THz radiation. This limitation can be mitigated in several aspects. For example, one can use semiconductor materials with higher mobility(e.g. low-temperature grown gallium arsenide, LT-GaAs) as well as short carrier lifetime (which is essential for THz generation) and increase the bias field as high as possible\citep{hou2013}. However, the amount of available materials is so small that there is not enough space for choice. In addition, the highest biased field is limited by the breakdown of the material. Attempt has also been made by designing micro-structures to generate singular and localized bias field within the active region of the PCA\citep{brener1996}. Moreover, line excitation of the laser beam has been used to make use of the trap-enhanced field near the anode and reduce the screening effect\citep{kim2005}. Most recently, electrodes with nano-structures are fabricated on the PCA to optimize the distribution of the incoming laser beam and improve the collection efficiency of the electrode\citep{park2012,heshmat2012,singh2013,tanoto2013,berry2013}. In this way, the enhancement of THz power as large as 50 times has been achieved\citep{berry2013}.

In this work, the design of a sawtooth-structured PCA is proposed, and its performance is compared with that of the conventional strip-line PCA by numerical simulation. Owing to these micro-structures, the localized bias fields are enhanced at the apex of the teeth. In addition, the elliptical other than circular laser beam is used for excitation in order to overlap with the enhanced fields near the apexes. The elliptical illumination can mitigate the screening effect caused by the space charge\citep{rodriguez1996}. Thereby, more photo-excited carriers can arrive at the electrode before recombination, and enhanced THz radiation can be obtained. Moreover, the elliptical illumination can also decrease laser's power density in the illuminated region, so that higher laser power can be applied before breakdown of the material occurs.

\section{Design of the sawtooth PCA}
\begin{figure}[h!]
\centering
\includegraphics[width=.8\textwidth]{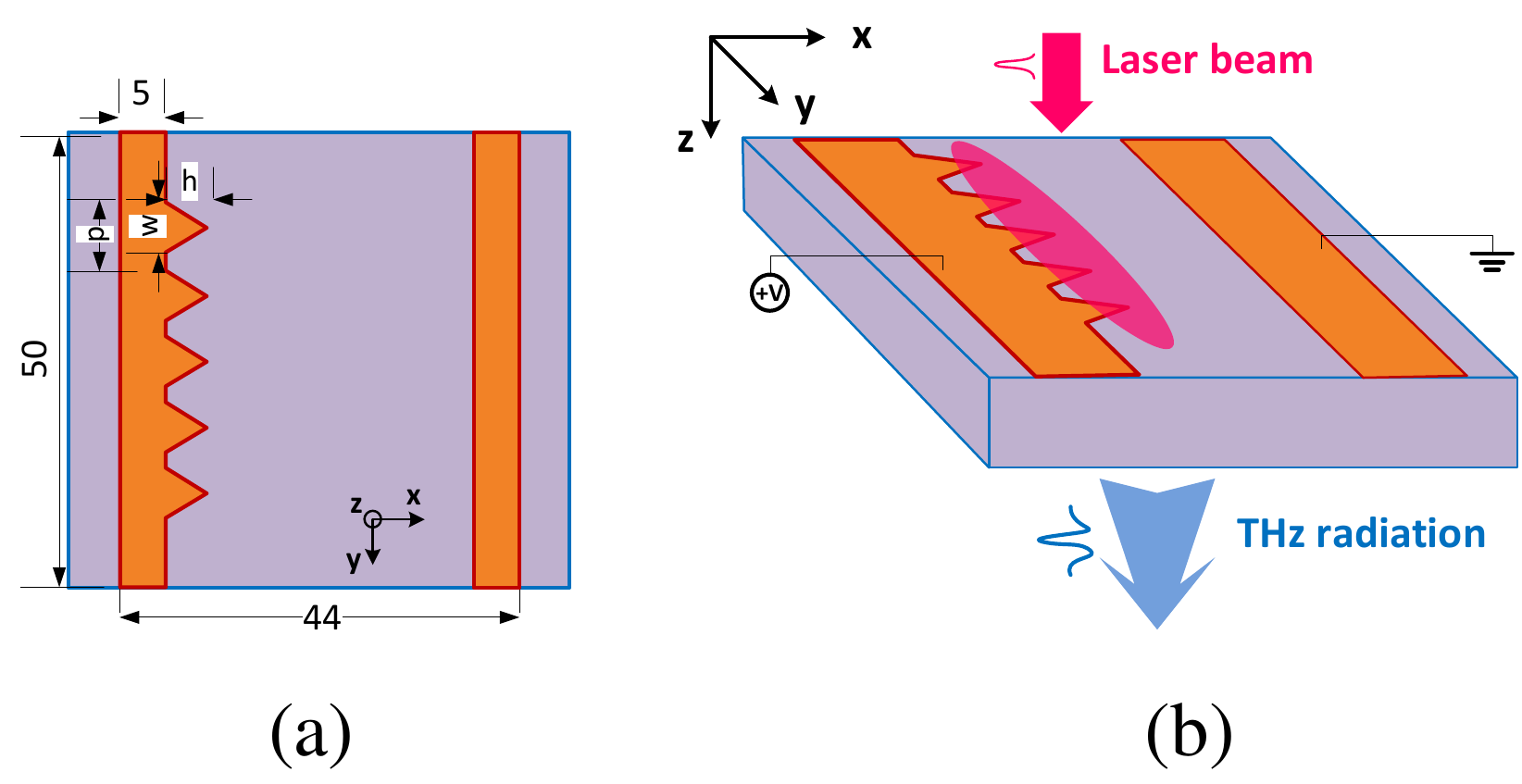}
\caption{Schematic structure and dimension of the sawtooth PCA. All dimensions are in the unit of $\mu m$.(a) The tooth has a height of $h=5.6\mu m$, width of $w=5.4\mu m$, and pitch size of $p=7\mu m$. (b) During the operation of the PCA, the incoming laser beam is partially overlapped with the sawtooth structure, and  the electrodes are externally biased by DC power supply. }
\label{fig_sawtooth_dimen}
\end{figure}

The dimension of the sawtooth PCA is shown in Fig. \ref{fig_sawtooth_dimen}. The total size is $44\mu m$ by $50\mu m$, and the width of the electrodes is $5\mu m$, with a gap of $34\mu m$ in between. The sawtooth structure is placed at the inner side of the anode. To generate THz radiation, the elliptical laser beam is applied for excitation, and the forward THz radiation in free space is collected at far-field.

\section{Numerical simulation result}

In our previous work\citep{jitao2014}, a full-wave simulation tool based on finite-difference-time-domain method has been developed to characterize and predicate the radiation properties of the PCA. The feature of this tool is that the multi-physical phenomena happening in the PCA, such as light-matter interaction, photo-excited carrier dynamics and full-wave propagation of the THz radiation, are considered and embodied in the simulation. Therefore, this tool can fulfill the needs of the comprehensive simulation of the PCA, in which almost all of the parameters that tightly related to the performance of the PCA can be involved, and the THz radiation can be predicated both in the near-field and far-field.

The performance of the sawtooth PCA is simulated by the full-wave tool. As a comparison, the conventional strip-line (without sawtooth structure) PCA with similar dimension is also simulated. The substrate of both PCAs is LT-GaAs with the thickness of $2.2\mu m$. For sawtooth PCA, the distance between the inner edge of the anode and the center of the elliptical beam is $12\mu m$. For strip-line PCA, circular beam with diameter of $20\mu m$ is used, and the distance between the inner edge of the anode and the center of the beam is $10\mu m$. Other parameters used in the simulation are summarized in Table.\ref{inputdata}.

\begin{figure}[H]
\centering
\includegraphics[width=1\textwidth]{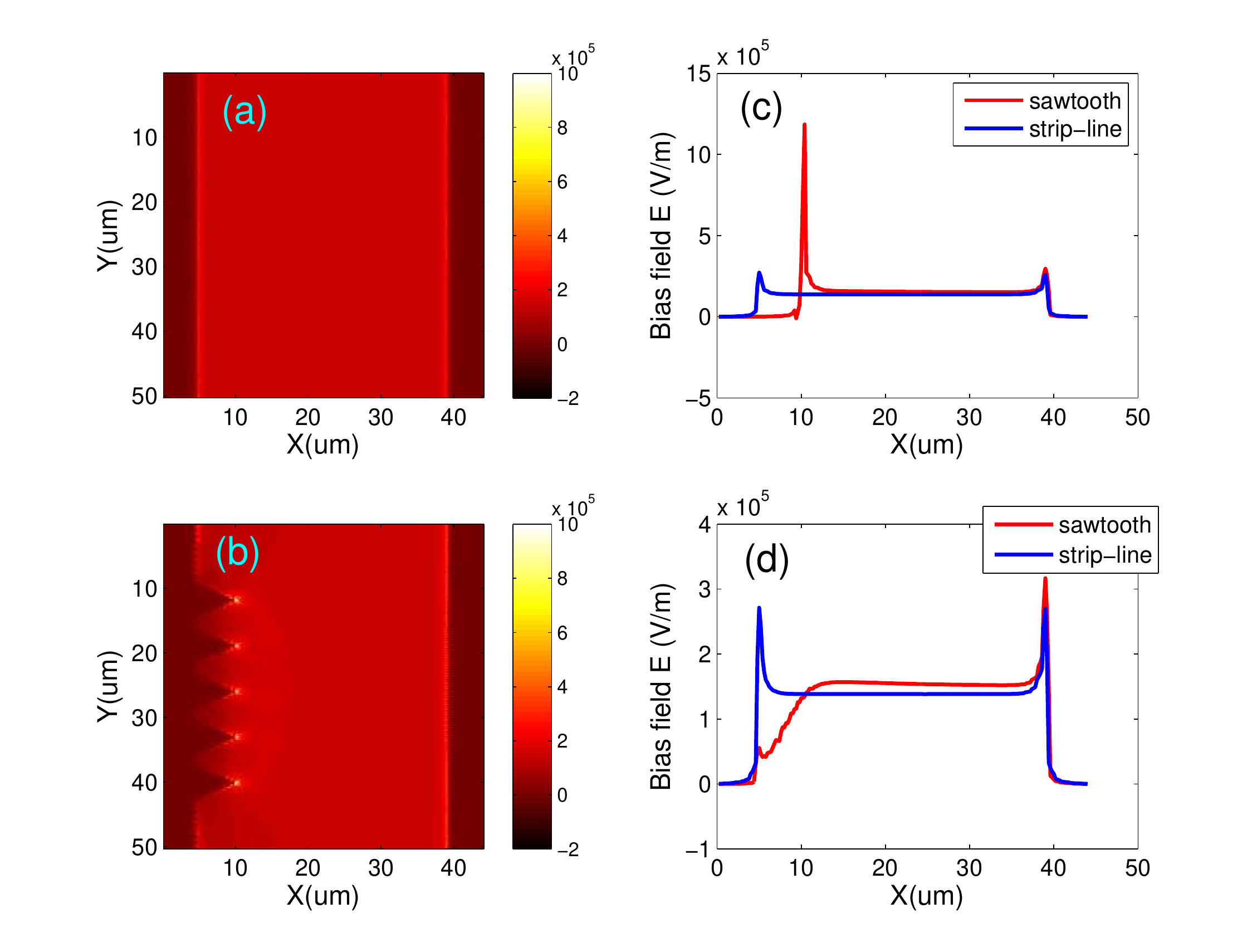}
\caption{Bias field Distribution of the sawtooth and strip-line PCAs. Only the field in x-axis is considered. Two-dimensional distribution of the bias field of (a) strip-line PCA and (b)sawtooth PCA at $XOY$ plane and $z=0$. Bias field comparison between strip-line and sawtooth PCAs along x-axis at (c)$y=26\mu m$ and (d)$y=29\mu m$.}
\label{fig_dc_sawtooth}
\end{figure}

One of predicted advantages of the sawtooth PCA is the enhanced bias field near the apex of the teeth. We first simulate the distribution of the DC field when the PCA is externally biased. The result is shown in Fig.\ref{fig_dc_sawtooth}. It can be seen from Fig.2(b) that the bias field near the apex of the teeth is much larger than other places. This field enhancement in sawtooth PCA is much clear when comparing with the strip-line PCA, as shown in Fig.2(c) \&(d).

\begin{table}[H]
\caption{Parameters used in the simulation}
\centering
    \begin{tabular}{ll}
    \hline
    \textbf{Parameters}                        & \textbf{Values}                   \\ \hline
    Material                          & LT-GaAs                  \\
    Carrier lifetime ($ps$)             & electron: 0.1, hole: 0.4 \\
    Mobility ($cm^{2}/V\cdot s$) & electron: 200, hole: 30  \\
    Permittivity                      & 12.9                     \\
    Intrinsic concentration ($cm^{-3}$) & 2.1E6                      \\
    Absorption coefficient ($cm^{-1}$) & 1E4                     \\
    Laser wavelength ($nm$)             & 800                      \\
    Beam size ($\mu m$)       & $20\times40$                  \\
    Pulse duration ($fs$)               & 80                       \\
    DC voltage ($V$)           & 5                      \\ \hline
    \end{tabular}
\label{inputdata}
\end{table}


\begin{figure}[H]
\centering
\includegraphics[width=1\textwidth]{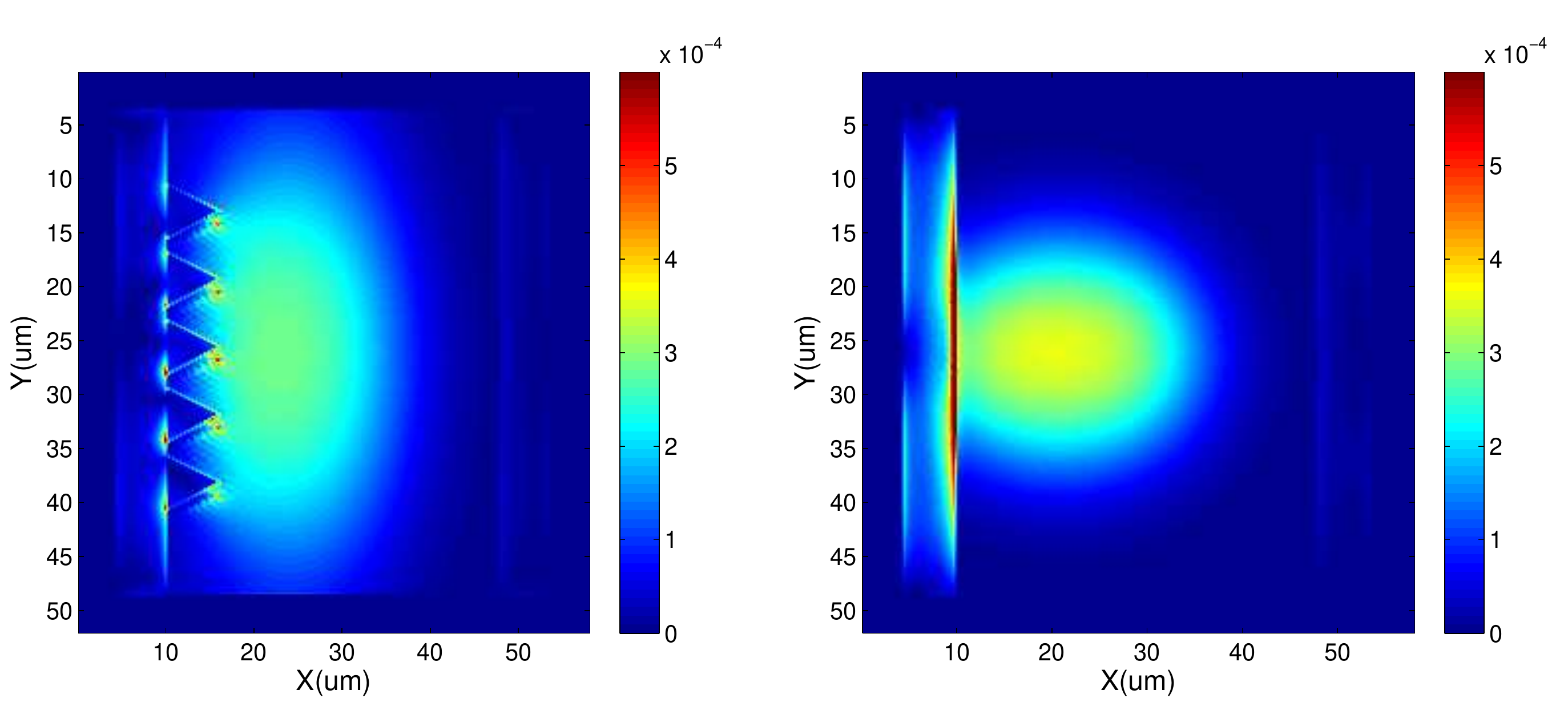}
\caption{Transient current density at the top surface of the (a) sawtooth PCA and (b) strip-line PCA. The colorbar has a unit of $A/\mu m^2$.}
\label{fig_current}
\end{figure}

When excited by the incoming laser pulse, the migration of the photo-excited carriers will result in current within the substrate. Due to the different structures, the distribution of the current in the sawtooth PCA and strip-line PCA is also different. Figure.\ref{fig_current} shows the transient current density at the top surface (i.e. $XOY$ plane) of the PCAs at the time of the laser pulse's peak ($0.5ps$). The mean laser power for both is $60mW$. In sawtooth PCA, the current is localized at the apex of the teeth. While in strip-line PCA, the current is distributed much uniformly along the edge of the electrode.

\begin{figure}[H]
\centering
\includegraphics[width=0.8\textwidth]{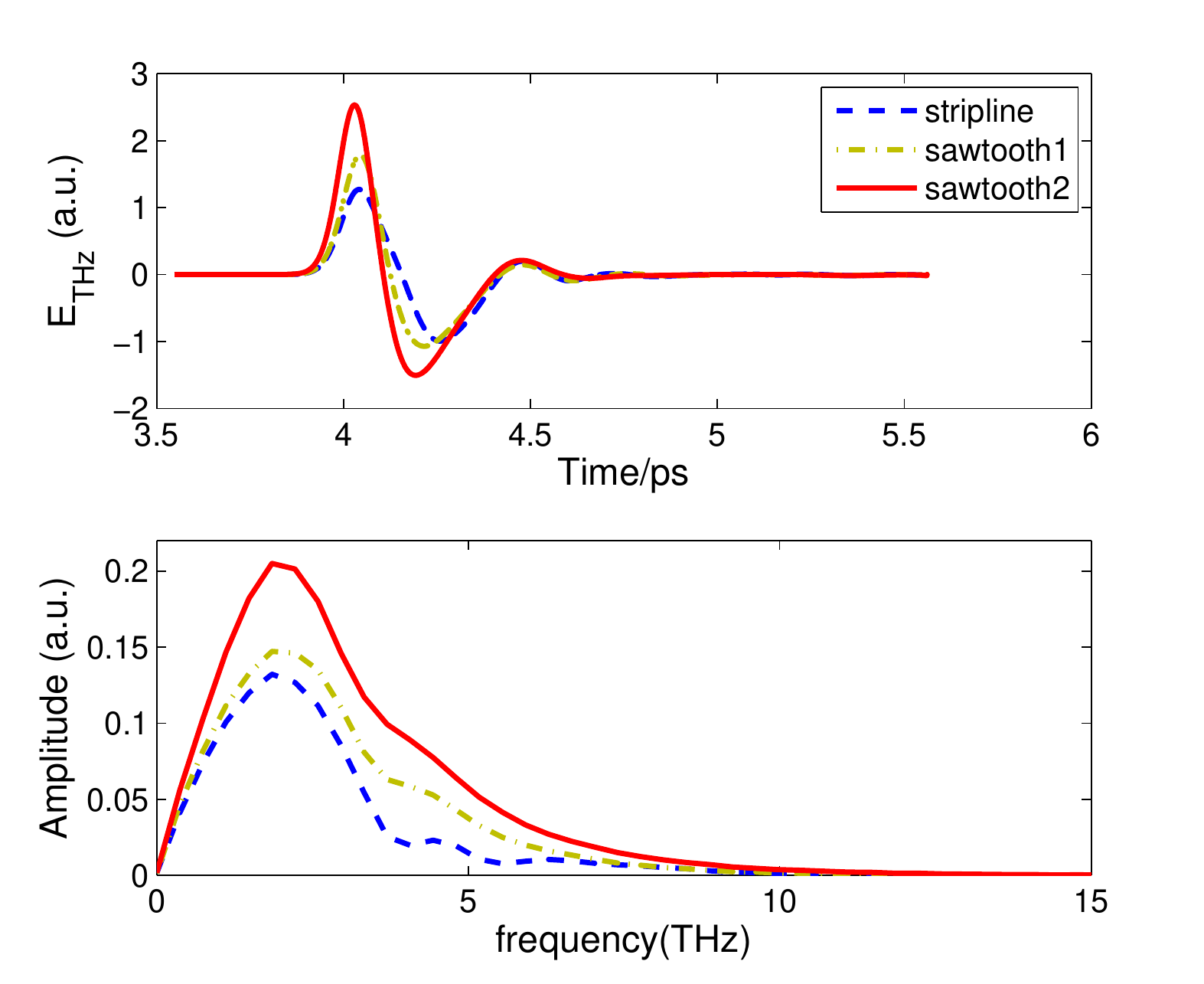}
\caption{Comparison of the far-field THz radiation between sawtooth PCA and strip-line PCA. (upper)time-domain THz pulse,and (lower)corresponding spectra by means of fast Fourier transformation. The yellow dot dish line and the red solid represent the THz field of the sawtooth PCA at the same laser power and the same intensity, respectively.}
\label{fig_far_field_60mw}
\end{figure}

To estimate the far-field radiation, the direction right below the PCA is chose and the distance for the center of the PCA is set as $200mm$. Since the THz radiation is polarized along $x$-axis , only the component of the field in this direction is calculated. The far-field radiation of both PCAs is compared at the same laser powers. Figure\ref{fig_far_field_60mw} shows the result when the laser power is $60mW$. It can be seen that the THz field(unless otherwise stated below, it refers to the peak of the time-domain THz pulse) of the sawtooth PCA is almost $1.4$ times of the strip-line PCA.


Moreover, for a common PCA, the material of the substrate usually suffers from the breakdown when the power density of the incoming laser beam is too much high. Since the elliptical beam illuminates larger area than the circular beam, more laser power can be applied to the PCA with elliptical-beam-illumination. To demonstrate this, we keep the peak intensity of the laser beam the same for sawtooth and strip-line PCAs (in this case the mean power of elliptical beam is twice of the circular beam ), and the corresponding far-field radiation is also shown in Fig.\ref{fig_far_field_60mw}. It is found that the THz radiation field of the sawtooth PCA is $2$ times of the strip-line PCA.

\section{Conclusion}
The design of a sawtooth PCA with elliptical-beam-illumination is proposed. Enhanced THz radiation is predicated by numerical analysis. The enhancement comes from the localized bias field near the apex of the teeth as well as the spread of the laser beam, which not only mitigates the screening effect but also enables the laser beam to illuminate the PCA with higher laser power before breakdown of the material happens. Further enhancement is promising by increasing the amount of the teeth along the electrode.

\section*{Acknowledgments}
This work can not be done without the contribution of the following colleagues. They are Mingguang Tuo, Min Liang, Xiong Wang and Hao Xin. They will be listed as co-authors when we consider to publish this work in a journal in future.


\bibliographystyle{plain}
\bibliography{references}
\end{document}